# Fluid-structure coupling of concentric double FGM shells with different lengths

Ehsan Moshkelgosha[1], Ehsan Askari[*2], Kyeong-Hoon Jeong[3] and Ali Akbar Shafiee[4]

[1] *Research Department, Samin Sanaat Shaigan Company, Isfahan, Iran*
[2] *CMEMS - Center for Microelectromechanical Systems, University of Minho, Azurém, 4800-058 Guimarães, Portugal*
[3] *Mechanical Engineering Division, Korea Atomic Energy Research Institute, 989-111 Daedeok-daero, Yuseong, Daejeon 305-353, Republic of Korea*
[4] *Mechanical Engineering Department, Isfahan University of Technology, Isfahan, Iran*

**Abstract.** The aim of this study is to develop a semi-analytical method to investigate fluid-structure coupling of concentric double shells with different lengths and elastic behaviours. Co-axial shells constitute a cylindrical circular container and a baffle submerged inside the stored fluid. The container shell is made of functionally graded materials with mechanical properties changing through its thickness continuously. The baffle made of steel is fixed along its top edge and submerged inside fluid such that its lower edge freely moves. The developed approach is verified using a commercial finite element computer code. Although the model is presented for a specific case in the present work, it can be generalized to investigate coupling of shell-plate structures via fluid. It is shown that the coupling between concentric shells occurs only when they vibrate in a same circumferential mode number, n. It is also revealed that the normalized vibration amplitude of the inner shell is about the same as that of the outer shell, for narrower radial gaps. Moreover, the natural frequencies of the fluid-coupled system gradually decrease and converge to the certain values as the gradient index increases.

## 1. Introduction

Functionally graded materials (FGMs) are advanced composites, microscopically engineered to have a smooth spatial variation of material properties in order to improve overall performance. Typically, FGMs are composed from a mixture of metals and ceramics and are further characterized by a smooth and continuous change of the mechanical properties from one surface to another. FGMs are regarded as one of the most promising candidates for future intelligent composites in many engineering fields such as heat exchanger tubes, biomedical implants, flywheels, blades, storage tanks, pressure vessels, and general wear and corrosion resistant coatings or for joining dissimilar materials in aerospace and automobile industries.

As the dynamic parameters play an important role in the design of modern advanced structures, many valuable studies on dynamic characteristics of inhomogeneous structures and in particular FGM cylindrical shells could be found in the literature. Moreover, dynamic behaviour of structures coupled with fluid is of great importance in various scientific and engineering applications, such as fluid-storage tanks, fuel tanks of space vehicles, nuclear reactors, and tower-like structures. Chen *et al*. (2004) considered the free vibration of simply supported, fluid-filled functionally graded cylindrical orthotropic shells with arbitrary thickness. Hasheminejad and Rajabi (2007) developed an exact methodology under the umbrella of the inherent background coefficients to consider the scattering of time-harmonic plane acoustic waves by a thick hollow isotropic FG cylinder immersed in and filled with non-viscous compressible fluids. Sheng and Wang (2008) reported the results of an investigation on the vibration of functionally graded cylindrical shells with a flowing fluid, embedded in an elastic medium, under mechanical and thermal loads. The vibration and stability of freely supported FGM truncated and complete conical shells subjected to uniform lateral and hydrostatic pressures were investigated by Sofiyev (2009). Iqbal *et al*. (2009) investigated the vibration characteristics of a functionally graded material circular cylindrical shell filled with fluid using a wave propagation approach. Sofiyev (2010) presented an analytical study on the dynamic behavior of the infinitely-long, FGM cylindrical shell subjected to combined action of the axial tension, internal compressive load and ring-shaped compressive pressure. The characteristics of beam-mode stability of fluid-conveying shell systems were investigated by Shen *et al*. (2014) for shells with boundary conditions, using a FEM algorithm. Wali *et al*. (2015) studied free vibration response of FGM shells by using an efficient three-dimensional shell model. Sofiyev and Kuruoglu (2015) studied the dynamic instability of three-layered cylindrical shells containing a functionally graded (FG) interlayer subjected to static and time dependent periodic axial compressive loads.

In addition, obstacles like baffles are usually used as damping devices to suppress the fluid sloshing motion in fluid-structure systems. Sloshing is a potential source of disturbance in fluid storage containers. The fluid in containers displays a free-surface fluctuation when the container is subjected to an external excitation, like an earthquake. This fluid sloshing may be a direct or indirect cause of the unexpected instability and failure in some engineering applications (Wang *et al*. 2016, Zhou *et al*. 2014, Askari *et al*. 2013, Askari *et al*. 2010). Some examples are the vibration analysis of fluid-storage tanks under earthquake waves, nuclear fuel storage pool, biomechanical systems, fuel storage tanks of aerospace vehicles and cargo tanks of LNG. To suppress sloshing, some internal bodies or baffles are generally placed in fluid storage tanks. It is well known that using such baffles results in energy dissipation and reduction in sloshing amplitude and hydrodynamic loads. In designing baffled containers, considering the sloshing phenomenon and the structural vibration are essential. It can be noted that baffled containers have two groups of modes: sloshing and bulging ones. Sloshing modes are caused by the oscillation of fluid free surface, whereas bulging modes are related to vibrations of the structure. In fact, it is well known that the effect of the free surface waves is low on bulging modes for structures that are not extremely flexible (Morand and Ohayon 1995, Askari and Daneshmand 2010, Askari and Daneshmand 2009). In the present paper, attention is focused on the bulging modes of baffled containers.

Cho *et al*. (2002) presented the fluid-structure interaction problems of fluid-storage containers with baffle by the structural-acoustic finite element formulation. Biswal *et al*. (2004) developed a finite element code and investigated the influence of a baffle on the dynamic response of a partially fluid-filled cylindrical tank. Gavrilyuk *et al*. (2006) presented fundamental solutions of the linearized problem on fluid sloshing in a vertical cylindrical baffled container. A pressure-based finite element technique has been developed to analyse the slosh dynamics of a partially filled rigid container with bottom-mounted submerged components by Mitra and Sinhamahapatra (2007). Biswal and Bhattacharyya (2010) employed finite element method to investigate the influence of composite baffles on the coupled dynamics of containers. Askari *et al*. (2011) proposed a semi-analytical method to study the effect of rigid internal bodies on partially fluid-filled containers. Wang *et al*. (2012, 2013) studied the effect of multiple rigid baffles on free and force vibration of a rigid cylindrical fluid-storage tank. Ebrahimian *et al*. (2014) developed a numerical model based on the boundary element method to determine an equivalent mechanical model for fluid storage containers with multiple baffles.

Although many remarkable studies have previously been conducted on dynamic analyses of shells made of functionally-graded materials, fluid-coupled vibration of FGM containers with baffles has not been investigated yet, based on the best knowledge of the authors. Moreover, investigating the influence of baffle flexibility on vibration of partially fluid-filled container, and vice versa, is of paramount importance in primary design stages of baffled containers due to the coupled vibration of structures through fluid.

Thus, this study considers the dynamic analysis of a partially fluid-filled cylindrical container with a flexible baffle, acquiring coupled modes and eigen-frequencies. The container is made of functionally graded materials with mechanical properties changing through its thickness continuously. A semi-analytical method is developed taking fluid-structure interaction into account. The fluid region is divided into two regions, namely the inner region and the outer region on the basis of the radius of the inner shell. The velocity potential of both fluid regions is described in terms of the Bessel functions and the sinusoidal functions appropriate to the boundary condition between the two distinct fluid regions. Using the collocation method, the Galerkin method and the superposition principle, relationships between the unknown coefficients of the velocity potential and the modal function of the shells are obtained as matrix forms. Finally, the Rayleigh-Ritz method is used to derive the frequency equation of the system. The developed methodology is applied to specific cases of fluid-coupled systems and verified comparing with a finite element analysis.

## 2. Mathematical formulation

Consider a baffle of length $h$, radius $b$, and uniform thickness $t_2$ submerged inside a cylindrical container of length $L$, radius $a$, and uniform thickness $t_1$, as it is depicted in Fig. 1. The baffle is fixed at its top and its lower edge freely moves in fluid domain. Meanwhile, the bottom of the container is rigid and flat, whereas its lateral wall is flexible. The container is assumed to be clamped-free and contains a fluid with a depth $H$, having a free surface. Moreover, the radial, circumferential and axial coordinates are illustrated by $r$, $\theta$ and $x$, respectively. The container is made of functionally graded materials and its mechanical properties change continuously according to gradually varying the volume fraction of the constituent materials, usually in the thickness direction (Hosseini-Hashemi *et al*. 2010). The fluid-storage tank is made of a composition of ceramic and metal with varying substance from its outer surface to inner surface, respectively. The modulus of elasticity and density vary linearly through the shell thickness according to a power-law distribution as

$$E(z) = (E_c - E_a)V_b(z) + E_a, \quad (1)$$

$$\rho(z) = (\rho_c - \rho_a)V_b(z) + \rho_a, \quad (2)$$

in which the subscripts $a$ and $c$ represent metallic and ceramic constituents, respectively, and the volume fraction $V_b$ may be given by (Hosseini-Hashemi *et al*. 2010, Shafiee *et al*. 2014)

$$V_b(z) = \left(\frac{z}{t} + \frac{1}{2}\right)^\alpha \quad (3)$$

where $\alpha$ is the gradient index and takes only a positive value. $z$ is measured from the middle surface of the shell towards outside, $-t/2 \leq z \leq t/2$. However, the Poisson's ratio is taken 0.3 over simulation. Typical values for metal and

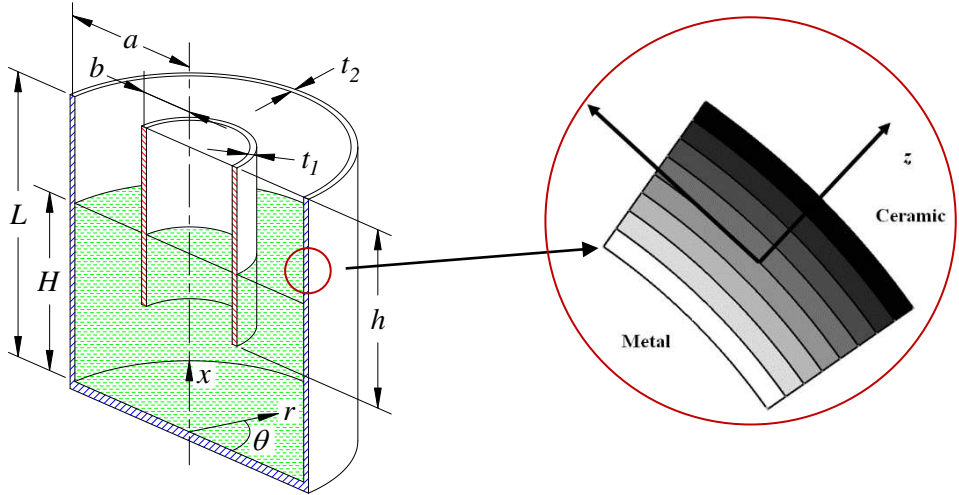

Fig. 1 a schematic representation of a partially fluid-filled cylindrical baffled container paired with a 2D slice of the container wall divided into 8 layers and made from a combination of ceramic and metal varying along the thickness direction ceramics used in the present article are listed in Table 1. Moreover, the fluid is water with a mass density $\rho_L$, which is assumed to be inviscid and incompressible.

Table 1 Material properties used in the FG shells

| Properties | Metal | Ceramic | |
|---|---|---|---|
| | Aluminum (Al) | Zirconia (ZrO$_2$) | Aluminia (Al$_2$O$_3$) |
| $E$ (GPa) | 70 | 200 | 380 |
| $\rho$ (kg/m$^3$) | 2702 | 5700 | 3800 |

*2.1 Rayleigh-Ritz method*

In order to obtain the characteristic equations of a dynamic system, the Rayleigh-Ritz method can be used by introducing the Rayleigh quotient (Zhu, 1995), for the dynamic system presented in this study, which is written as

$$\omega^2 = \frac{U_{S_1} + U_{S_2}}{T^*_{S_1} + T^*_{S_2} + T^*_L} \quad (4)$$

where $U_{S_1}$ and $U_{S_2}$ are the potential energies associated with the container and baffle, respectively. $T^*_{S_1}$ and $T^*_{S_2}$ also are the reference kinetic energies of the shells, while $T^*_L$ is the simplified fluid reference kinetic energy due to shell movements. Finally, $\omega$ is the natural frequency of the fluid-coupled system.

Knowing that the material properties of a cylindrical shell made of functionally graded materials change continuously through its thickness, the general potential energy of the shell, according to the assumption of shell theory, is as follows (Leissa 1969, Soedel 2003)

$$U = \iint_\Omega [\frac{K}{2}(\varepsilon_x^2 + \varepsilon_\theta^2) + K_{x\theta}(\varepsilon_x \varepsilon_\theta) + \frac{K_G}{2}\varepsilon_{x\theta}^2] R d\Theta$$
$$+ \iint_\Omega [\frac{D}{2}(\kappa_x^2 + \kappa_\theta^2) + D_{x\theta}\kappa_x\kappa_\theta + \frac{D_G}{2}\kappa_{x\theta}^2] R d\Theta]$$
$$+ \iint_\Omega [C(\varepsilon_x \kappa_x + \varepsilon_\theta \kappa_\theta) + C_{x\theta}(\varepsilon_x \kappa_\theta + \varepsilon_\theta \kappa_x)$$
$$+ C_G \varepsilon_{x\theta} \kappa_{x\theta}) R d\Theta], \quad d\Theta = dx\,d\theta \quad (5)$$

where the middle surface of the shell is delineated as $\Omega$ in Eq. (5), and the membrane stiffness is defined as

$$K = \int_{-t/2}^{t/2} \frac{E(z)}{1-\vartheta^2(z)} dz, \quad (6)$$

$$K_{x\theta} = \int_{-t/2}^{t/2} \frac{E(z)\vartheta(z)}{1-\vartheta^2(z)} dz, \quad (7)$$

$$K_G = \int_{-t/2}^{t/2} G(z) dz, \quad (8)$$

the bending stiffness

$$D = \int_{-t/2}^{t/2} \frac{E(z)z^2}{1-\vartheta^2(z)} dz, \quad (9)$$

$$D_{x\theta} = \int_{-t/2}^{t/2} \frac{E(z)\vartheta(z)z^2}{1-\vartheta^2(z)} dz, \quad (10)$$

$$D_G = \int_{-t/2}^{t/2} G(z) z^2 dz, \quad (11)$$

and the membrane-bending coupled stiffness

$$C = \int_{-t/2}^{t/2} \frac{E(z)z}{1-\vartheta^2(z)} dz, \quad (12)$$

$$C_{x\theta} = \int_{-t/2}^{t/2} \frac{E(z)\vartheta(z)z}{1-\vartheta^2(z)} dz, \quad (13)$$

$$C_G = \int_{-t/2}^{t/2} G(z) z dz, \quad (14)$$

Furthermore, the strain-displacement relations of the FGM cylindrical shell in the cylindrical coordinates are the same as homogeneous ones, the strains and curvatures of the shell at the its middle surface are as follows

$$\varepsilon_x = \frac{\partial u}{\partial x}, \quad (15)$$

$$\varepsilon_\theta = \frac{1}{a}(\frac{\partial^2 w}{\partial \theta^2} + w), \tag{16}$$

$$\varepsilon_{x\theta} = \frac{1}{a}\frac{\partial u}{\partial \theta} + \frac{\partial v}{\partial x}, \tag{17}$$

$$\kappa_x = -\frac{\partial^2 w}{\partial x^2}, \tag{18}$$

$$\kappa_\theta = -\frac{1}{a^2}(\frac{\partial^2 w}{\partial \theta^2} - \frac{\partial v}{\partial \theta}), \tag{19}$$

$$\kappa_{x\theta} = -\frac{1}{a}(2\frac{\partial^2 w}{\partial x \partial \theta} - \frac{\partial v}{\partial x}), \tag{20}$$

where $u$, $v$, and $w$ are the axial, circumferential and radial displacements of the shell on its middle surface, respectively. The reference kinetic energy of the FGM shell, as discussed by Soedel (2003), is given by

$$T^* = \frac{1}{2}\tilde{\rho}\iint_\Omega \dot{q}\cdot\dot{q}d\Omega = \frac{1}{2}\tilde{\rho}\iint_\Omega [u^2(x,\theta)+v^2(x,\theta)+w^2(x,\theta)]d\Omega \tag{21}$$

where $\tilde{\rho}$ is the surface mass density written as follows

$$\tilde{\rho} = \int_{-t/2}^{t/2}\rho(z)dz, \tag{22}$$

To apply the Rayleigh-Ritz method, the modal functions, (Blevins 1979), of the container shell with any boundary conditions are written as a linear combination of admissible functions

$$u_1(x,\theta) = \sum_{n=1}^{\infty}\sum_{m=1}^{\infty}U_{mn}\frac{1}{\eta_m}\frac{\partial \psi_m(x)}{\partial x}\cos(n\theta),$$
$$\eta_m = \frac{\beta_m}{h} \tag{23}$$

$$v_1(x,\theta) = \sum_{n=1}^{\infty}\sum_{m=1}^{\infty}V_{mn}\psi_m(x)\sin(n\theta) \tag{24}$$

$$w_1(x,\theta) = \sum_{n=1}^{\infty}\sum_{m=1}^{\infty}W_{mn}\psi_m(x)\cos(n\theta) \tag{25}$$

where $U_{mn}$, $V_{mn}$ and $W_{mn}$ are the parameters of the Rayleigh-Ritz expansion; the symbols $n$ and $m$ are the circumferential wave number and the axial mode number, respectively. The dry beam function, $\psi_m(x)$, is used as the admissible function satisfying imposed boundary conditions and $\beta_m$ can be calculated from the characteristic equations based on boundary conditions, as discussed by Blevins (1979). The modal functions for the cylindrical baffle are also given by

$$u_2(x,\theta) = \sum_{s=1}^{\infty}\sum_{k=1}^{\infty}\tilde{U}_{ks}\frac{1}{\chi_k}\frac{\partial \Lambda_k(x)}{\partial x}\cos(s\theta), \chi_k = \frac{\phi_k}{L} \tag{26}$$

$$v_2(x,\theta) = \sum_{s=1}^{\infty}\sum_{k=1}^{\infty}\tilde{V}_{ks}\Lambda_k(x)\sin(s\theta) \tag{27}$$

$$w_2(x,\theta) = \sum_{s=1}^{\infty}\sum_{k=1}^{\infty}\tilde{W}_{ks}\Lambda_k(x)\cos(s\theta) \tag{28}$$

where $\tilde{U}_{ks}$, $\tilde{V}_{ks}$ and $\tilde{W}_{ks}$ are parameters of the Rayleigh-Ritz expansion. The symbols, $s$ and $k$ indicate the circumferential wave number and the axial mode number, respectively. The dry beam function $\Lambda_k(x)$ is used as the admissible function, satisfying imposed boundary conditions and $\phi_k$ can be calculated from the characteristic equations (Blevins 1979). After extracting the potential energies and kinetic energies of cylindrical structures, the only remaining term to be computed in Eq. (4) is the simplified fluid kinetic energy due to shell movements, $T_L^*$, which can be written by using the Green's theorem (Amabili *et al.* 1998)

$$T_L^* = \frac{1}{2}\rho_L\omega^2\iint_S \varphi\frac{\partial\varphi}{\partial\varsigma}dS \tag{29}$$

where $S = S_2^I + S_2^{II} + S_I$ that $S_2^I$ and $S_2^{II}$ are the lateral wet surface of the baffle in the regions (*I*) and (*II*), respectively. The symbol, $S_1$, is the lateral wet surface of the cylindrical container. $\varsigma$ also is the normal direction to the surface $S$ of the fluid domain. Furthermore, the variable, $\varphi$ in Eq. (29) is the fluid displacement potential.

### 2.2 Velocity and displacement potentials

The inviscid, incompressible and irrotational fluid permits the introduction of a velocity potential for the fluid motion (Eftekhari 2016). The liquid velocity potential can be expressed as

$$\tilde{\varphi}(r,\theta,x,t) = i\omega\varphi(r,\theta,x)e^{i\omega t}, \quad i^2 = -1. \tag{30}$$

where the fluid displacement potential $\varphi$ satisfies the superposition principle stated in Eq. (31), (Amabili *et al.* 1998).

$$\varphi = \varphi_1 + \varphi_2 \tag{31}$$

The fluid displacement potential associated with the motion of the cylindrical container is denoted by $\varphi_1$, while the baffle is assumed to be rigid. Conversely, $\varphi_2$ is the fluid displacement potential in conjunction with the motion of the baffle when the container is considered stationary. In order to formulate fluid motion, the fluid domain can be assumed to consist of two separated parts, regions (*I*) and (*II*) as follows

$$\begin{aligned}&\text{Region } I = \{(r,\theta,x): \ r<b, \ x<H\}\\&\text{Region } II = \{(r,\theta,x): \ b<r<a, \ x<H\}.\end{aligned} \tag{32}$$

That is to say, the region (*I*) is the cylindrical inner liquid domain, and the region (*II*) the annular outer liquid domain. Three interfacing surfaces between the shells and fluid are defined as shown in Eq. (33) and Fig. 1.

$$\gamma = \{(r,\theta,x): \ r = b, 0 < \theta < 2\pi, x < L - h\}$$
$$\gamma' = \{(r,\theta,x): \ r = b, 0 < \theta < 2\pi, L - h < x < H\} \quad (33)$$
$$\gamma'' = \{(r,\theta,x): \ r = a, 0 < \theta < 2\pi, 0 < x < H\}$$

where the symbol $\gamma$ represents the interface between two fluid regions along the axial gap between the inner shell and the container bottom. In addition, $\gamma'$ is the interfacing boundary between the fluid and the inner shell along either region (*I*) or region (*II*), while $\gamma''$ is the fluid-interfacing boundary of the outer shell in region (*II*).

*2.2.1 Fluid displacement potential associated with the container, $\varphi_1$*

The spatial displacement potential, $\varphi_1$, can be written for both fluid regions (Askari *et al.* 2013).

$$\varphi_1^I = \sum_{m=1}^{\infty}\sum_{n=1}^{\infty} W_{mn}\varphi_{mn}^I, \quad \varphi_1^{II} = \sum_{m=1}^{\infty}\sum_{n=1}^{\infty} W_{mn}\varphi_{mn}^{II}, \quad (34)$$

where $\varphi_{mn}^I$ and $\varphi_{mn}^{II}$ are the displacement potential functions in the fluid regions (*I*) and (*II*) induced by the vibration of the outer shell, respectively. The modal parameter $W_{mn}$ is used for the Rayleigh-Ritz expansion defined in Eq. (25). The displacement potential should satisfy not only the Laplace equation of Eq. (35), but also boundary conditions and compatibility equation of Eqs. (36)-(39).

$$\nabla^2 \varphi_1 = 0, \quad (35)$$

$$\left.\frac{\partial \varphi_1}{\partial r}\right|_{r=b} = 0, \quad \text{along} \quad L - h < x < H, \quad (36)$$

$$\left.\frac{\partial \varphi_1}{\partial x}\right|_{x=0} = 0, \quad (37)$$

$$\left.\varphi_1\right|_{x=H} = 0, \quad (38)$$

$$\left.\frac{\partial \varphi_1}{\partial r}\right|_{r=a} = w_1 \quad \text{along} \quad 0 < x < H, \quad (39)$$

The problem defined by the above relations allows the separation of spatial variables in the cylindrical coordinate system for each liquid region

$$\varphi_{in}^I = \cos(n\theta)\sum_{m=1}^{\infty} A'_{mni} \cos(\lambda_m x) I_n(\lambda_m r),$$
$$\lambda_m = \frac{\pi(2m-1)}{2H}, \quad (40)$$

and

$$\varphi_{in}^{II} = \cos(n\theta)\sum_{m=1}^{\infty} \cos(\lambda_m x) \times [B'_{mni} I_n(\lambda_m r) + C'_{mni} K_n(\lambda_m r)] \quad (41)$$

where $A'_{mni}$, $B'_{mni}$ and $C'_{mni}$ are unknown coefficients depending on integers *m*, *n* and *i*. $I_n$ and $K_n$ are the modified Bessel functions of the first and second kind of order *n*, respectively. Using Eq. (39) and integrating the resultant relation term by term between 0 and *H*, the coefficient, $C'_{mni}$ can be written as a function of the unknown coefficient, $B'_{mni}$, by the well known properties of orthogonal trigonometric functions.

$$C'_{mni} = \frac{1}{K'_n(\lambda_m a)}\left\{\frac{2}{H}\int_0^H \psi_i(x)\cos(\lambda_m x)\,dx - B'_{mni} I'_n(\lambda_m a)\right\} \quad (42)$$

where $I'_n$ and $K'_n$ indicate the derivatives of $I_n$ and $K_n$, respectively. Moreover, the compatibility conditions and the boundary conditions, Eqs. (35)-(39), must be so satisfied as to be found the unknown coefficients contained in Eqs. (43)-(45).

$$\left.\frac{\partial \varphi_1^I}{\partial r}\right|_{r=b} = \begin{cases}\left.\dfrac{\partial \varphi_1^{II}}{\partial r}\right|_{r=b} & \text{along } \gamma \\ 0 & \text{along } \gamma'\end{cases} \quad (43)$$

$$\varphi_1^{II} = \varphi_1^I \quad \text{along } \gamma, \quad (44)$$

$$\left.\frac{\partial \varphi_1^{II}}{\partial r}\right|_{r=b} = 0 \quad \text{along } \gamma'. \quad (45)$$

Solving the above equations, simultaneously by employing the Galerkin method leads to Eq. (46)

$$\begin{cases}\Theta_1 \mathbf{A}'_{ni} = \Theta_2 \mathbf{B}'_{ni} + \Theta_3 \\ \mathbf{P}_1 \mathbf{A}'_{ni} = \mathbf{P}_2 \mathbf{B}'_{ni} + \mathbf{P}_3\end{cases} \quad (46)$$

in which the unknown coefficients matrices, $\mathbf{A}'_{ni}$ and $\mathbf{B}'_{ni}$, are given in Appendix A.

*2.2.2 Fluid displacement potential associated with the baffle, $\varphi_2$*

The fluid displacement potential for each fluid region is assumed to be

$$\varphi_2^I = \sum_{k=1}^{\infty}\sum_{s=1}^{\infty} \tilde{W}_{ks}\Omega_{ks}^I, \quad \varphi_2^{II} = \sum_{k=1}^{\infty}\sum_{s=1}^{\infty} \tilde{W}_{ks}\Omega_{ks}^{II} \quad (47)$$

where $\tilde{W}_{ks}$ is the parameters of the Ritz expansion defined in Eq. (28). $\Omega_{ks}^I$ and $\Omega_{ks}^{II}$ are displacement potential eigen-functions in liquid regions (*I*) and (*II*), respectively. The displacement liquid potential should satisfy the Laplace equation given in Eq. (48) and boundary conditions of Eqs. (49)-(52).

$$\nabla^2 \varphi_2 = 0, \quad (48)$$

$$\left.\frac{\partial \varphi_2}{\partial r}\right|_{r=b} = w_2, \quad \text{along} \quad L - h < x < H \quad (49)$$

$$\left.\frac{\partial \varphi_2}{\partial x}\right|_{x=0} = 0 \quad (50)$$

$$\left.\varphi_2\right|_{x=H} = 0 \quad (51)$$

$$\left.\frac{\partial \varphi_2}{\partial r}\right|_{r=a} = 0, \quad \text{along} \quad 0 < x < H \quad (52)$$

In the inner liquid region (*I*), the separation of the

variables gives

$$\Omega_{ks}^{I} = \sum_{f=1}^{\infty} \cos(s\theta) A_{fsk} \cos(\lambda_f x) I_s(\lambda_f r),$$
$$\lambda_f = \frac{\pi(2f-1)}{2H}$$
(53)

and in the outer liquid region (*II*), it gives

$$\Omega_{ks}^{II} = \sum_{f=1}^{\infty} \cos(s\theta) B_{fsk} \cos(\lambda_f x)$$
$$\times \left[ I_s(\lambda_f r) - \frac{I'_s(\lambda_f a)}{K'_s(\lambda_f a)} K_s(\lambda_f r) \right]$$
(54)

where $A_{fsk}$ and $B_{fsk}$ are unknown coefficients. It also is necessary to ensure that fluid displacement potentials and dynamic displacement normal to wet surfaces are continuous along $\gamma$, and Eq. (49) should be satisfied along $\gamma'$ (Askari *et al.* 2011). These requirements are listed in Eqs. (55)-(57).

$$\left.\frac{\partial \varphi_2^I}{\partial r}\right|_{r=b} = \begin{cases} \left.\dfrac{\partial \varphi_2^{II}}{\partial r}\right|_{r=b} & \text{on } \gamma \\ w_2 & \text{on } \gamma' \end{cases}$$
(55)

$$\varphi_2^{II} = \varphi_2^{II} \qquad \text{on } \gamma$$
(56)

$$\left.\frac{\partial \varphi_2^{II}}{\partial r}\right|_{r=b} = w_2 \qquad \text{on } \gamma'$$
(57)

Substituting Eqs. (53) and (54) into Eqs. (55)-(57) and employing the collocation method, the unknown coefficients, $A_{fsk}$ and $B_{fsk}$ can be written as a matrix form of Eq. (58).

$$\begin{cases} \mathbf{Q}_1 \mathbf{A}_{sk} = \mathbf{Q}_2 \mathbf{B}_{sk} + \mathbf{Q}_3 \\ \mathbf{R}_1 \mathbf{A}_{sk} = \mathbf{R}_2 \mathbf{B}_{sk} + \mathbf{R}_3 \end{cases}$$
(58)

Matrices existing in Eq. (58) are defined in Appendix B.

### 2.3 Reference fluid kinetic energy

The total kinetic energy of fluid can be obtained expanding Eq. (29), which is a summation of the kinetic energies of liquid as shown in Eq. (59).

$$T_L^* = \frac{1}{2}\rho_L$$
$$\times \int_0^{2\pi}\int_{L-h}^H \left[(\varphi_1^I + \varphi_2^I)w_2 - (\varphi_1^{II} + \varphi_2^{II})w_2\right]_{r=b} b\,\mathrm{d}x\,\mathrm{d}\theta$$
$$+ \frac{1}{2}\rho_L \int_0^{2\pi}\int_0^H \left(\varphi_1^{(II)} + \varphi_2^{(II)}\right)_{r=a} w_1 a\,dx\,d =$$
$$T_L^{(2)} + T_L^{(1-2)} + T_L^{(1)} + T_L^{(2-1)}$$
(59)

where $T_L^{(1)}$ and $T_L^{(2)}$ are the kinetic energy terms owing to the interaction between the outer and inner shells with liquid, respectively. In addition, the kinetic energy terms, $T_L^{(1-2)}$ and $T_L^{(2-1)}$ indicate the interaction between the shells via liquid (Amabili *et al.* 1998).

Although there are no contacts between the shells, they are interactive through fluid, which is called as the coupling effect of fluid. In fact, two terms of $T_L^{(1-2)}$ and $T_L^{(2-1)}$ in Eq. (59) are the reference kinetic energy terms associated with the interaction between the shells via fluid.

$$T_L^{(1-2)} = \frac{1}{2}\rho_L \int_0^{2\pi}\int_{L-h}^H \left[(\varphi_1^I - \varphi_1^{II})\right]_{r=b} w_2 b\,\mathrm{d}x\,\mathrm{d}\theta$$
(60)

$$T_L^{(2-1)} = \frac{1}{2}\rho_L \int_0^{2\pi}\int_0^H \left(\varphi_2^{(II)}\right)_{r=a} w_1 a\,dx\,d\theta$$
(61)

so

$$T_L^{(1-2)} = \frac{1}{2}\rho_L b \times \sum_{n=1}^{\infty}\sum_{s=1}^{\infty}\sum_{i=1}^{\infty}\sum_{j=1}^{\infty}\sum_{m=1}^{\infty} W_{in}\widetilde{W}_{js}$$
$$\times \left[ -B_{mni}\left\{I_n(\lambda_m b) - \frac{I'_n(\lambda_m a)}{K'_n(\lambda_m a)}K_n(\lambda_m b)\right\} + \frac{2}{H}\frac{K_n(\lambda_m b)}{K'_n(\lambda_m a)}\int_0^H \psi_i(x)\cos(\lambda_m x)\,\mathrm{d}x \right.$$
$$+ A'_{mni}I_n(\lambda_m b)\right] \times$$
$$\int_{L-h}^H \cos(\lambda_m x)\Lambda_j(x)dx \int_0^{2\pi}\cos(n\theta)\cos(s\theta)d\theta$$
$$= \frac{1}{2}\mathbf{W}\,\mathbf{OS}_1\,\widetilde{\mathbf{W}}$$
(62)

and

$$T_L^{(2-1)} = \frac{1}{2}\rho_L a \sum_{n=1}^{\infty}\sum_{s=1}^{\infty}\sum_{f=1}^{\infty}\sum_{i=1}^{\infty}\sum_{j=1}^{\infty} \widetilde{W}_{is} W_{jn} B_{fsi}$$
$$\times \left[(I_s(\lambda_f a) - \frac{I'_s}{K'_s}(\lambda_f a)K_s(\lambda_f a))\right] \times$$
$$\int_0^{2\pi}\cos(n\theta)\cos(s\theta)d\theta \int_0^H \cos(\lambda_s x)\psi_j(x)dx$$
$$= \frac{1}{2}\widetilde{\mathbf{W}}\,\mathbf{OS}_2\,\mathbf{W}$$
(63)

Moreover, the term representing the interaction between the liquid and outer shell in Eq. (44) is

$$T_L^{(1)} = \frac{1}{2}\rho_L \int_0^{2\pi}\int_0^H \left(\varphi_1^{II}\right)_{r=a} w_1 a\,dx\,d\theta$$
$$= \frac{1}{2}\rho_L a \sum_{n=1}^{\infty}\sum_{s=1}^{\infty}\sum_{i=1}^{\infty}\sum_{j=1}^{\infty}\sum_{m=1}^{\infty} W_{in}W_{js}$$
$$\times \left[B_{mni}\left\{I_n(\lambda_m a) - \frac{I'_n(\lambda_m a)}{K'_n(\lambda_m a)}K_n(\lambda_m a)\right\}\right.$$
$$+ \frac{2}{H}\frac{K_n(\lambda_m a)}{K'_n(\lambda_m a)}\int_0^H \psi_i(x)\cos(\lambda_m x)\,\mathrm{d}x\right] \times$$
$$\int_0^H \cos(\lambda_m x)\psi_j(x)dx \int_0^{2\pi}\cos(n\theta)\cos(s\theta)d\theta$$
$$= \frac{1}{2}\mathbf{W}\,\mathbf{OP}_1\,\mathbf{W}$$
(64)

and the interaction between the inner shell and fluid is the term $T_L^{(2)}$, which is

$$T_L^{(2)} = \frac{1}{2}\rho_L \int_0^{2\pi}\int_{L-h}^H \left[(\varphi_2^I - \varphi_2^{II})\right]_{r=b} w_2 b\,\mathrm{d}x\,\mathrm{d}\theta$$
$$= \frac{1}{2}\rho_L b \sum_{n=1}^{\infty}\sum_{s=1}^{\infty}\sum_{i=1}^{\infty}\sum_{j=1}^{\infty}\sum_{f=1}^{\infty} \widetilde{W}_{in}\widetilde{W}_{js}$$

$$\left[-B_{fsi}\left\{I_s(\lambda_f b) - \frac{I'_s}{K'_s}(\lambda_f a)K_s(\lambda_f b)\right\}\right.$$
$$\left. + A_{fsi}I_s(\lambda_f b)\right]$$
$$\times \int_{L-h}^{H} \cos(\lambda_f x)\Theta_j(x)dx \int_0^{2\pi} \cos(n\theta)\cos(s\theta)d\theta \qquad (65)$$
$$= \frac{1}{2}\tilde{\mathbf{W}} \mathbf{OP}_2 \tilde{\mathbf{W}}$$

### 2.4 Eigenvalue extraction

For the numerical calculation of natural frequencies and Ritz unknown coefficients, the limited expansion terms of admissible modal functions should be considered. So, $N$ is the number of expansion terms in the circumferential admissible modal functions for $s$ and $n$. On the other hand, the number $M$ indicates the number of expansion terms in the axial admissible modal functions for $m$ and $k$. The number of expansion terms $N$ and $M$ should be chosen large enough to give required accuracy. The vector $\mathbf{Q}$ of parameters in the Ritz expansion is defined as

$$\mathbf{Q} = \begin{Bmatrix} \mathbf{q} \\ \tilde{\mathbf{q}} \end{Bmatrix}, \qquad \mathbf{q} = \begin{Bmatrix} \mathbf{U} \\ \mathbf{V} \\ \mathbf{W} \end{Bmatrix}, \qquad \tilde{\mathbf{q}} = \begin{Bmatrix} \tilde{\mathbf{U}} \\ \tilde{\mathbf{V}} \\ \tilde{\mathbf{W}} \end{Bmatrix} \qquad (66)$$

where $\mathbf{q}$ and $\tilde{\mathbf{q}}$ are the displacement vectors of the container and baffle shells, respectively. The displacement vectors are defined as follows

$$\mathbf{U} = \begin{Bmatrix} U_{(1,1)} \\ \vdots \\ U_{(1,M)} \\ U_{(2,1)} \\ \vdots \\ U_{(2,M)} \\ \vdots \\ U_{(N,1)} \\ \vdots \\ U_{(N,M)} \end{Bmatrix}, \mathbf{V} = \begin{Bmatrix} V_{(1,1)} \\ \vdots \\ V_{(1,M)} \\ V_{(2,1)} \\ \vdots \\ V_{(2,M)} \\ \vdots \\ V_{(N,1)} \\ \vdots \\ V_{(N,M)} \end{Bmatrix}, \mathbf{U} = \begin{Bmatrix} W_{(1,1)} \\ \vdots \\ W_{(1,M)} \\ W_{(2,1)} \\ \vdots \\ W_{(2,M)} \\ \vdots \\ W_{(N,1)} \\ \vdots \\ W_{(N,M)} \end{Bmatrix} \qquad (67)$$

and similarly

$$\tilde{\mathbf{U}} = \begin{Bmatrix} \tilde{U}_{(1,1)} \\ \vdots \\ \tilde{U}_{(1,M)} \\ \tilde{U}_{(2,1)} \\ \vdots \\ \tilde{U}_{(2,M)} \\ \vdots \\ \tilde{U}_{(N,1)} \\ \vdots \\ \tilde{U}_{(N,M)} \end{Bmatrix}, \tilde{\mathbf{V}} = \begin{Bmatrix} \tilde{V}_{(1,1)} \\ \vdots \\ \tilde{V}_{(1,M)} \\ \tilde{V}_{(2,1)} \\ \vdots \\ \tilde{V}_{(2,M)} \\ \vdots \\ \tilde{V}_{(N,1)} \\ \vdots \\ \tilde{V}_{(N,M)} \end{Bmatrix}, \tilde{\mathbf{U}} = \begin{Bmatrix} \tilde{W}_{(1,1)} \\ \vdots \\ \tilde{W}_{(1,M)} \\ \tilde{W}_{(2,1)} \\ \vdots \\ \tilde{W}_{(2,M)} \\ \vdots \\ \tilde{W}_{(N,1)} \\ \vdots \\ \tilde{W}_{(N,M)} \end{Bmatrix} \qquad (68)$$

The maximum potential energy of both of cylindrical shells, Eq. (5), becomes

$$U_S = U_{S1} + U_{S2} = \frac{\pi}{2}\mathbf{Q}^T\mathbf{K}\mathbf{Q}, \qquad (69)$$

with the partitioned matrix $\mathbf{K}$, which is

$$\mathbf{K} = \begin{bmatrix} [\mathbf{K}_{S1}] & [\mathbf{0}] \\ [\mathbf{0}] & [\mathbf{K}_{S2}] \end{bmatrix}, \qquad (70)$$

where $\mathbf{K}_{S1}$ and $\mathbf{K}_{S2}$ are the stiffness matrixes of the outer and inner cylindrical shells, respectively, which are defined in Appendix C. The symbol of $[\mathbf{0}]$ indicates a null matrix. The reference kinetic energy of both of the cylindrical shells, Eq. (21) can be rewritten as

$$T_S^* = T_{S1}^* + T_{S2}^* = \frac{\pi}{2}\mathbf{Q}^T\mathbf{M}\mathbf{Q}, \qquad (71)$$

where

$$\mathbf{M} = \begin{bmatrix} [\mathbf{M}_{S1}] & [\mathbf{0}] \\ [\mathbf{0}] & [\mathbf{M}_{S2}] \end{bmatrix}, \qquad (72)$$

The sub-matrices, $\mathbf{M}_{S1}$ and $\mathbf{M}_{S2}$ in Eq. (72) are the mass matrices of the outer and inner shells respectively, which are given in Appendix C. The simplified reference kinetic energy of fluid, Eq. (59), can be rewritten as

$$T_L^* = \frac{1}{2}\pi \mathbf{Q}^T \mathbf{M}_L \mathbf{Q}, \qquad (73)$$

where

$$\mathbf{M}_L = \begin{bmatrix} [\mathbf{0}] & [\mathbf{0}] & [\mathbf{0}] & [\mathbf{0}] & [\mathbf{0}] & [\mathbf{0}] \\ [\mathbf{0}] & [\mathbf{0}] & [\mathbf{0}] & [\mathbf{0}] & [\mathbf{0}] & [\mathbf{0}] \\ [\mathbf{0}] & [\mathbf{0}] & [\mathbf{OP}_1] & [\mathbf{0}] & [\mathbf{0}] & [\mathbf{OS}_1] \\ [\mathbf{0}] & [\mathbf{0}] & [\mathbf{0}] & [\mathbf{0}] & [\mathbf{0}] & [\mathbf{0}] \\ [\mathbf{0}] & [\mathbf{0}] & [\mathbf{0}] & [\mathbf{0}] & [\mathbf{0}] & [\mathbf{0}] \\ [\mathbf{0}] & [\mathbf{0}] & [\mathbf{OS}_2] & [\mathbf{0}] & [\mathbf{0}] & [\mathbf{OP}_2] \end{bmatrix}, \qquad (74)$$

Substituting Eqs. (69), (71) and (73) into the Rayleigh quotient, Eq. (4), and then minimizing it with respect to the coefficients $Q_i$, we obtain

$$\mathbf{KQ} - \omega^2(\mathbf{M} + \mathbf{M}_L)\mathbf{Q} = \mathbf{0}, \qquad (75)$$

which is a linear eigenvalue problem for a real, non-symmetric matrix.

### 2.5 Coupled modes

The coupling between the baffle and container occurs only when they are vibrating in a same circumferential mode number, $n$. It can be realized with taking a look at the elements of the stiffness and the mass matrices, $\mathbf{M}$, $\mathbf{M}_L$ and $\mathbf{K}$. This conclusion is very close to what was reported by Au-Yang (1975). It is worth noting that the greater the gap between the cylindrical shells is, the less they observe coupling effects as they vibrate in a same circumferential mode number. If the gap is large enough, the shells are not influenced by the hydrodynamic coupling. Once the shells are vibrating with different circumferential mode numbers, they vibrate independently. It means that they do not affect each other via fluid. Actually, it can be assumed that one shell vibrates whereas the other shell is not deformed.

## 3. Numerical result and discussion

In this section, the characteristic equation of a cylindrical fluid-storage tank with a flexible baffle are solved using a written MATLAB code. The material properties and geometry of the system are as follows; the baffle made of steel has Young's modulus $E$=206 GPa, Poisson's ratio of $\vartheta$=0.3, and mass density of $\rho_s$=7850 kg/m$^3$. The liquid as water has depth of $H$=0.4 m, mass density of $\rho_L$=1000 kg/m$^3$. Meanwhile, the container is made of functionally graded materials, such as Al/ZrO$_2$ and Al/Al$_2$O$_3$, and their material properties are indicated in Table 1. The baffle has a radius of $b$=0.1 m, length of $h$=0.45 m, and wall thickness of $t$=0.002 m. The axial gap between the baffle and container also is $d$=0.15 m. The radius of container with a thickness of $t$=0.002 m is assumed to be $a$=0.11 m or 0.2 m with a length of 0.6 m.

For verification purposes, a finite element (FE) model of the fluid-coupled system is also constructed and obtained results are compared to those acquired by the semi-analytical simulation presented in this study. In the FE analysis, the FGM container is modelled by considering a multi-layered composite as each layer has constant properties and a number of layers are chosen. They could therefore be simulated correctly by changing FGM mechanical properties through thickness direction. A commercial FEM code, ABAQUS (version 6.5) is used for numerical analysis. In order to assure a high precision for the finite element results, the container shell is divided into 8 layers, Fig. 1. The structural model is constructed with a quadrilateral shell element (S4R) which is a four-node, doubly curved shell element with a reduced integration, hourglass control, and the finite membrane strain formulation. The liquid region is meshed with acoustic elements (AC3D8), which are eight-node brick acoustic elements with a linear interpolation, and the element has only one unknown pressure per node (Virella *et al.* 2006). A density $\rho$=1000 kg/m$^3$ and a bulk modulus $K$=2.07 GPa, i.e., the properties of water, are used in the computations. Acoustic three-dimensional finite elements based on linear wave theory are used to represent the hydrodynamics of fluid. The location of each node on the constrained surfaces of liquid coincides exactly to the location of corresponding node of structure. Along the interface between liquid and structure, the fluid surface is tied to the shell surface in normal direction to satisfy compatibility conditions. This contact formulation is based on the master-slave approach, in which both surfaces remain in contact throughout simulation, allowing the transmission of normal forces between them. As no pressure was applied to nodes at the free fluid surface, no sloshing waves are considered in this study. In addition, the number of elements used in the finite element model is listed in Table 2.

Two investigations are delineated in the section. First of all, a functionally graded baffled container is taken into consideration. The baffle is assumed as a vertical cylindrical rigid shell, which is submerged into liquid. The effect of rigid inner shell on dynamic characteristics of the FGM flexible cylindrical tank is also estimated as a function of the baffle radius. The second considered case stresses on the dynamic analysis of a flexible steel-made baffle and a flexible FGM container with material properties listed in Table 1. The effect of gradient indices on the first and second natural frequencies of the fluid-coupled system for several circumferential modes is also investigated. Furthermore, the effect of the liquid depth on natural frequencies and the radius of the structures on the mode shapes of coupled system are studied. For all these investigations, the verifications of the theoretical natural frequencies of the coupled system using the FEM analyses are also provided.

Table 2 Total number of elements used in the finite element analysis ($b$=0.1 m)

| Model | $a$ | Number of elements | | |
|---|---|---|---|---|
| | | (S4R) | | (AC3D8) |
| | | Outer shell | Inner shell | fluid |
| A | 0.11 | 5800 | 4350 | 39600 |
| B | 0.2 | 7200 | 5400 | 56300 |

### 3.1 Case 1: Flexible FGM container with a rigid baffle

Obstacles like baffles have been usually used in moving containers to suppress free surface waves. Considering the effect of baffles on the hydroelastic vibration and the sloshing phenomenon of containers is of paramount importance during design process. The vibration modes of baffled containers can be categorized into two modes, sloshing modes and bulging modes. Sloshing modes are caused by the oscillation of the fluid free surface, whereas bulging modes are related to vibrations of the flexible structures coupled with the liquid. Moreover, it is well known that the free surface waves negligibly affect bulging modes of structures that are not extremely flexible. In the present study, our attention is focused on the bulging modes of the liquid-coupled system.

Tables 3 and 4 summarize natural frequencies of the fluid-coupled structures shown in Fig. 1. As it can be observed, the obtained results are in agreement with those of finite element analyses. In addition, it is illustrated that the rigid baffle with a radius of $a$=0.11 m affects natural frequencies of the tank much more than that with $a$=0.2 m. it means that the narrower gap between the two structures, the more reduction in the natural frequencies due to increasing the effect of fluid-structure interaction. Moreover, the fundamental modes of the system with these two different radii of the container, $a$=0.11 and 0.2 m, corresponds to the first mode with the circumferential wave number $n$=2 and 3, respectively.

The effect of the annular gap between the container and baffle on natural frequencies for the first and second axial modes of the system is plotted in Figs. 2 and 3 as a function of the radius of rigid baffle ($b$). It is illustrated that natural frequencies decrease as baffle radius increases with a sharp decrease as the gap is narrower regardless of mode numbers. Moreover, it can be observed that the influence of the baffle radius on natural frequencies is relatively significant for lower circumferential mode numbers due to the separation effect (Askari and Jeong 2010). An increase of circumferential modes produces nodal points along the periphery of shell, dividing the circumferential fluid flow near the shell. At the same time, it reduces the active vibrating surface of the shell. Consequently, the effect of the liquid inertia on system dynamics decreases, which is called a "separation effect".

Table 3 Natural frequencies of the coupled system with the FGM container, Al/ZrO$_2$

| Mode n,m | This study | FEM (ABAQUS) | This study | FEM (ABAQUS) |
|---|---|---|---|---|
| | a=0.11 | | a=0.2 | |
| 1,1 | 200.74 | 205.08 | 381.03 | 376.00 |
| 1,2 | ——— | ——— | ——— | ——— |
| 2,1 | 102.53 | 99.44 | 197.07 | 193.37 |
| 2,2 | 438.06 | 430.27 | 632.61 | 615.80 |
| 3,1 | 169.47 | 163.15 | 126.16 | 125.04 |
| 3,2 | 374.99 | 369.12 | 419.27 | 409.25 |
| 4,1 | 289.13 | 281.35 | 147.36 | 143.25 |
| 4,2 | 543.89 | 536.24 | 330.26 | 324.10 |
| 5,1 | 477.02 | 469.61 | 210.47 | 202.39 |
| 5,2 | 727.83 | 714.27 | 347.00 | 336.32 |

Table 4 Natural frequencies of the coupled system with the FGM container, Al/Al$_2$O$_3$

| Mode n,m | This study | FEM (ABAQUS) | This study | FEM (ABAQUS) |
|---|---|---|---|---|
| | a=0.11 | | a=0.2 | |
| 1,1 | 250.48 | 243.46 | 478.99 | 472.03 |
| 1,2 | ——— | ——— | ——— | ——— |
| 2,1 | 124.91 | 121.99 | 248.86 | 239.54 |
| 2,2 | 564.87 | 557.23 | 814.71 | 800.25 |
| 3,1 | 213.24 | 209.04 | 155.84 | 152.41 |
| 3,2 | 482.42 | 476.33 | 541.79 | 534.94 |
| 4,1 | 373.39 | 376.07 | 182.17 | 176.27 |
| 4,2 | 705.23 | 690.41 | 428.35 | 413.89 |
| 5,1 | 614.2 | 602.55 | 269.57 | 261.03 |
| 5,2 | ——— | ——— | 451.52 | 440.19 |

### 3.2 Case 2: Flexible FGM container with a flexible baffle

Usually ceramic tiles have been utilized to laminate the outer part of cylindrical containers, protecting the container from the erosion induced by an aggressive chemical environment. They also increase the thermal resistance of the container against a hight temperature of surroundings. However, these tiles are prone to crack and to debond the shell/tile interface due to an abrupt transition between thermal expansion coefficients. In other words, a difference in the thermal expansion of the materials causes stress concentrations at the interface of the ceramic tiles and the shell, which results in cracking or debonding. An FGM composed of the ceramics on the outside surface and the metal on the inside surface of the shell eliminates the abrupt transition between the thermal expansion coefficients, offers the thermal resistance and the corrosion protection, and increases the load carrying capability. This is possible because the material composition of an FGM changes gradually through the thickness. Therefore, stress concentrations due to an abrupt change in material properties can be eliminated.

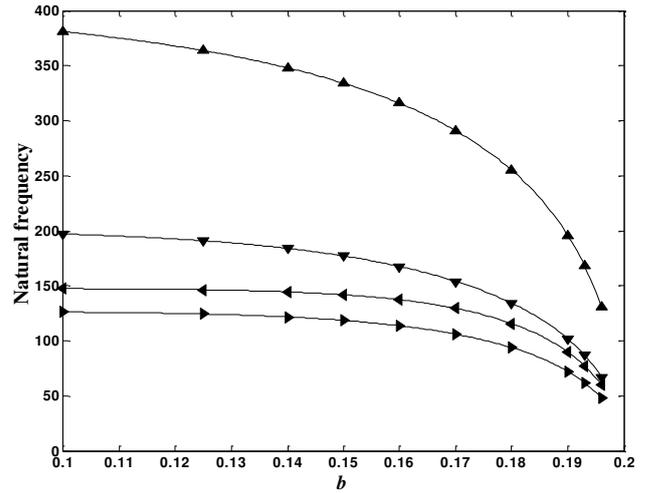

Fig. 2 Effect of annular gap on natural frequencies of the fluid-coupled system with the fundamental axial mode, $m=1$. ($H=0.4$ m) (—▲—, $n=1$; —▼—, $n=2$; —▶—, $n=3$; —◀—, $n=4$)

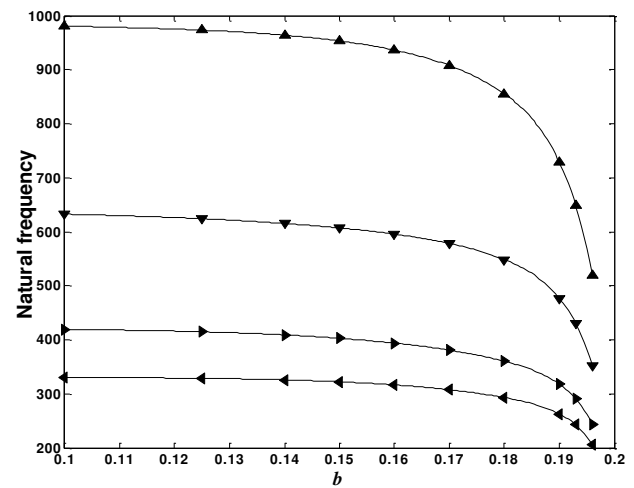

Fig. 3 Effect of the annular gap on natural frequencies of the fluid-coupled system with the second axial mode, $m=2$. ($H=0.4$ m) (—▲—, $n=1$; —▼—, $n=2$; —▶—, $n=3$; —◀—, $n=4$)

Regarding to industrial applications, the dynamic analyses of a baffled container made of FGMs were studied employing the model developed within this work, together with finite element method. Tables 5 and 6 listed natural frequencies of the coupled system for two sets of material properties, Al/ZrO$_2$ and Al/Al$_2$O$_3$, respectively, which illustrated small discrepancies, comparing outcomes acquired by semi-analytical method and finite element method. The effect of gradient indices on the two lowest natural frequencies was also investigated and illustrated in Figs. 4 and 5 for Al/Al$_2$O$_3$ and Figs. 6 and 7 for Al/ZrO$_2$. It could be observed from the figures that the natural frequencies gradually decrease and converge to the certain values as the gradient index increases. It can be explained that an increase of the gradient index indicates that the dominant material properties of the container shell changes from the ceramic to Aluminum. When the gradient index, $\alpha$ is 0, the mechanical properties of the shell converge the ceramics' properties, Al$_2$O$_3$ or ZrO$_2$. When the gradient index, $\alpha$ converges to infinite, the properties of the structure, however, approach that of Aluminium. Natural frequencies dominantly decrease as the gradient index is within the range of 0-5. It should be noted that the gradient indices affect the natural frequencies of Al/Al$_2$O$_3$ shells more than those of Al/ZrO$_2$ shells.

Table 5 Natural frequencies of the coupled system with the FGM container, Al/ZrO$_2$

| Mode n,m | This study | FEM (ABAQUS) | This study | FEM (ABAQUS) |
|---|---|---|---|---|
| | a=0.11 | | a=0.2 | |
| 1,1 | 144.93 | 142.01 | 312.29 | 317.52 |
| 1,2 | 444.59 | 432.77 | 437.48 | 422.68 |
| 2,1 | 74.764 | 72.80 | 140.74 | 135.79 |
| 2,2 | 170.66 | 165.62 | 202.65 | 197.48 |
| 3,1 | 118.27 | 118.00 | 125.74 | 123.66 |
| 3,2 | 244.74 | 236.04 | 194.12 | 198.65 |
| 4,1 | 228.84 | 217.08 | 147.21 | 141.70 |
| 4,2 | 394.48 | 378.22 | 329.4 | 315.96 |
| 5,1 | 401.55 | 390.37 | 210.32 | 202.15 |
| 5,2 | 575.51 | 554.95 | 346.86 | 337.01 |

Table 6 Natural frequencies of the coupled system with the FGM container, Al/Al$_2$O$_3$

| Mode n,m | This study | FEM (ABAQUS) | This study | FEM (ABAQUS) |
|---|---|---|---|---|
| | a=0.11 | | a=0.2 | |
| 1,1 | 160.48 | 157.06 | 328.53 | 332.05 |
| 1,2 | 515.1 | 506.23 | 524.2 | 513.94 |
| 2,1 | 80.472 | 77.34 | 141.83 | 137.14 |
| 2,2 | 196.17 | 189.67 | 254.2 | 249.00 |
| 3,1 | 126.89 | 129.60 | 154.73 | 158.20 |
| 3,2 | 285.58 | 277.82 | 194.96 | 189.85 |
| 4,1 | 255.72 | 248.45 | 181.94 | 174.37 |
| 4,2 | 433.8 | 425.02 | 376.04 | 368.22 |
| 5,1 | 456.61 | 444.76 | 269.35 | 260.17 |
| 5,2 | 633.65 | 621.88 | 451.29 | 443.76 |

Figs. 8-11 show the first four mode shapes of liquid-coupled baffled container with radii of 0.11 and 0.2 m for n=3 and 4. It is worth mentioning that the normalized amplitude of the baffle is about the same as that of the container as the annular gap is a=0.11, showing a strong interaction between two concentric shells, Figs. 8 and 9. Furthermore, it can be observed some modes are the distinct in-phases and the others are the distinct out-of-phases. However, the normalized vibration amplitude of the baffle

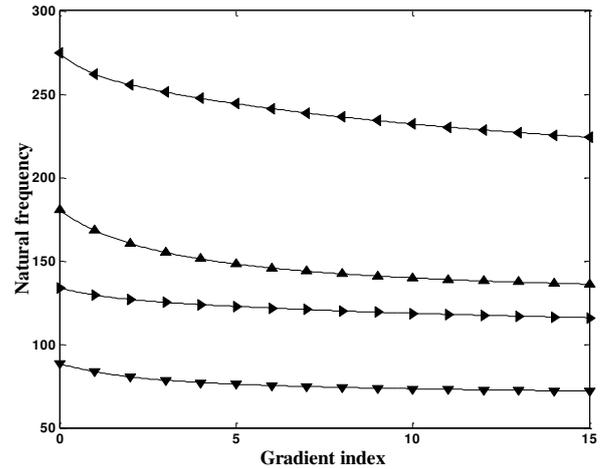

Fig. 4 Gradient index ($\alpha$) effect with the material Al/Al$_2$O$_3$ on the natural frequencies of the fluid-coupled system for the first axial mode, $m$=1. ($b$=0.1 m) (◀— , $n$=1; ▼— , $n$=2; ▶— , $n$=3; ◀— , $n$=4)

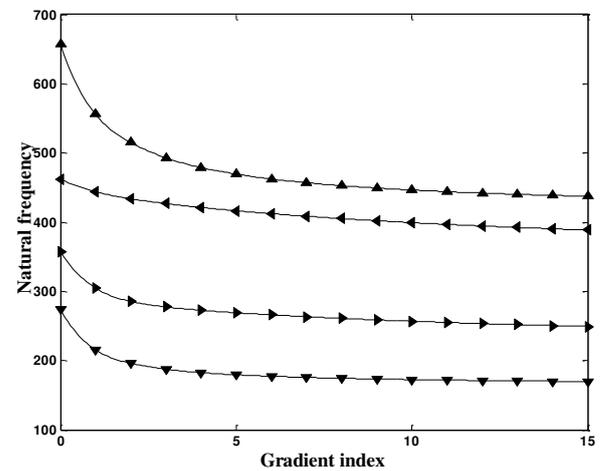

Fig. 5 Gradient index ($\alpha$) effect with the material Al/Al$_2$O$_3$ on the natural frequencies of the fluid-coupled system for the second axial mode, $m$=1. ($b$=0.1 m) (▲— , $n$=1; ▼— , $n$=2; ▶— , $n$=3; ◀— , $n$=4)

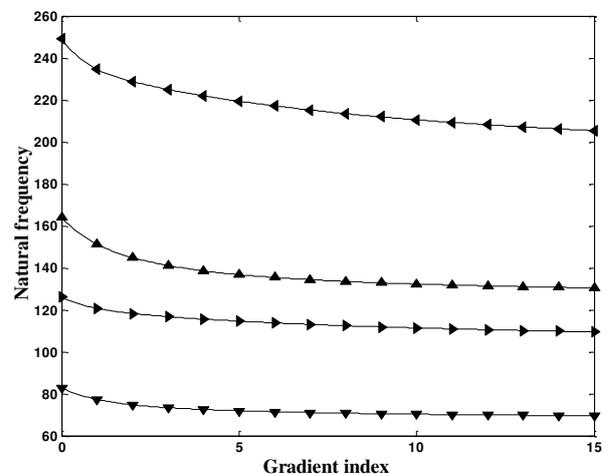

Fig. 6 Gradient index ($\alpha$) effect with the material Al/ZrO$_2$ on the natural frequencies of the fluid-coupled system for the first axial mode, $m$=1. ($b$=0.1 m) (▲— , $n$=1; ▼— , $n$=2; ▶— , $n$=3; ◀— , $n$=4)

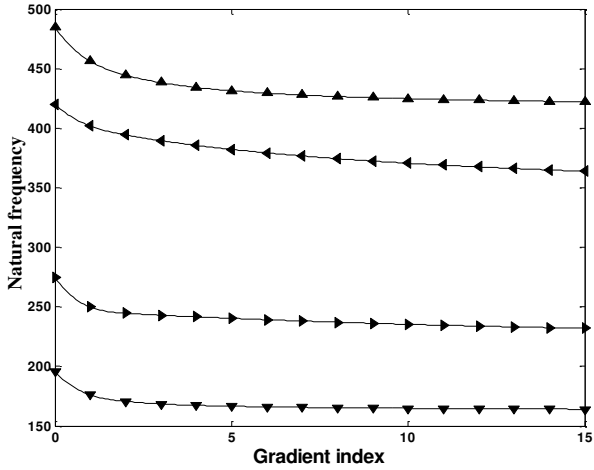

Fig. 7 Gradient index (*α*) effect with the material Al/ZrO$_2$ on the natural frequencies of the fluid-coupled system for the second axial mode, *m*=1. (*b*=0.1 m) (▲, *n*=1; ▼, *n*=2; ▶, *n*=3; ◀, *n*=4)

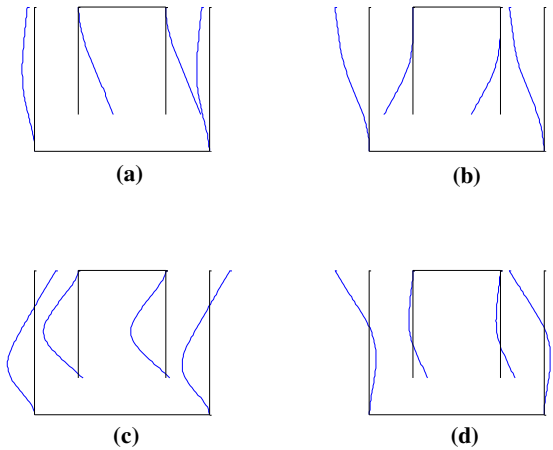

Fig. 8 First four mode shapes of the baffled container for *n*=3 and *a*=0.11

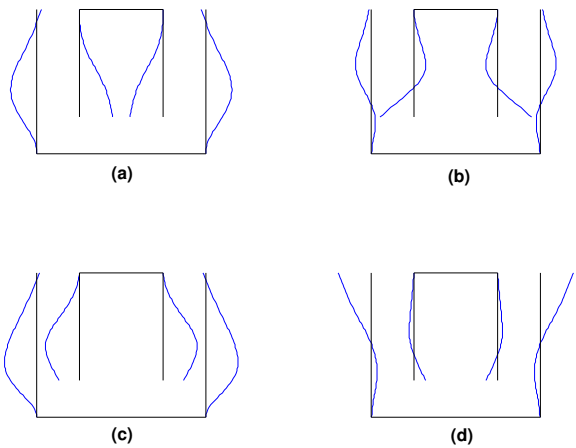

Fig. 9 First four mode shapes of the baffled container for *n*=4, *a*=0.11

differs from that of the container, for an annular gap of *a*=0.20, due to a weak interaction between the structures via the liquid, Figs. 10 and 11. That is to say, the shells move

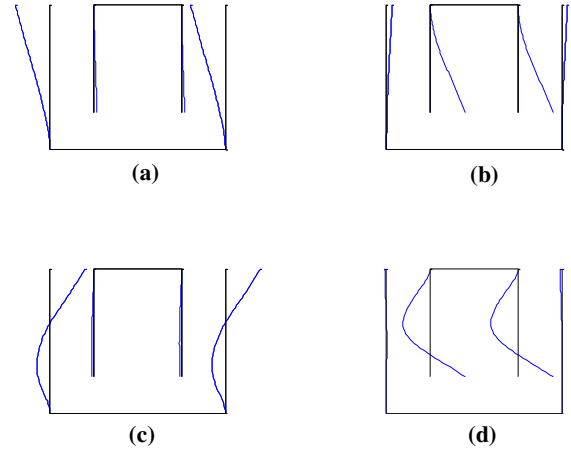

Fig. 10 First four mode shapes of the baffled container for *n*=3 and *a*=0.2

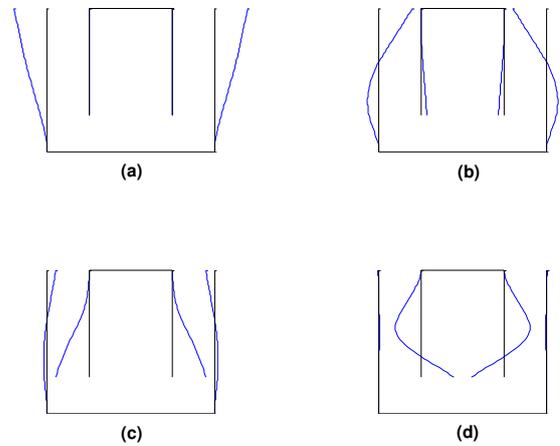

Fig. 11 First four mode shapes of the baffled container for *n*=4 and *a*=0.2

independently comparing with the case with the narrower gap.

## 4. Conclusions

A semi-analytical method to study the coupled dynamics of FGM containers with flexible baffles was developed taking fluid-structure interaction into account. Fundamental frequencies and modes of the system were acquired using the Rayleigh quotient, the Galerkin method, the collocation method and the superposition principle. The theoretical model enabled us to assess vibration of concentric shells coupled via fluid. Furthermore, the effect of functionally-graded materials on the fluid-coupled vibration of the system was evaluated employing this methodology.

It was concluded that fundamental frequencies of the coupled system decrease as the baffle radius increases, that is, the radial gap decreases. In particular, a sharp reduce in frequencies was observed as the radial gap became narrower regardless of mode numbers. In addition, the gradient index affected the system frequencies which decreased and converged to the certain values. The natural frequencies also decreased as the container fluid depth raised. The first four mode shapes of the fluid-coupled system with *a*=0.11 and 0.2 for *n*=3 and 4 were also presented. It was illustrated that the baffle and the container vibrated with about similar normalized amplitude for narrower gaps such as 0.01 m, showing a strong interaction between the shells. According to mathematical formulation and obtained results, it was also concluded that the shells oscillated independently as they vibrated with different circumferential mode numbers.

*CC*

**Appendix A:** the unknown matrices in Eq. (46)

In this appendix, the unknown matrices in Eq. (46) are given in detail. Unknown matrices in Eq. (A1) are as follows

$$\mathbf{\Theta}_1 \mathbf{A}'_{ni} = \mathbf{\Theta}_2 \mathbf{B}'_{ni} + \mathbf{\Theta}_3, \tag{A.1}$$

$$\Theta_1(j,m) = \frac{H}{2} I'_n(\lambda_m b) \delta_{jm}$$

$$\Theta_2(j,m) = \left[ I'_n(\lambda_m b) - \frac{I'_n(\lambda_m a)}{K'_n(\lambda_m a)} K'_n(\lambda_m b) \right] \times \hbar_{mj}$$

$$\Theta_3(j) = \sum_{m=1}^{\infty} \frac{2}{H} \frac{K'_n(\lambda_m b)}{K'_n(\lambda_m a)} \times \left( \int_0^H \psi_i(x) \cos(\lambda_m x) \, dx \right) \times \hbar_{mj} \tag{A.2}$$

$$\mathbf{A}'_{ni} = \begin{Bmatrix} A'_{1ni} \\ \vdots \\ A'_{mni} \end{Bmatrix}, \quad \mathbf{B}'_{ni} = \begin{Bmatrix} B'_{1ni} \\ \vdots \\ B'_{mni} \end{Bmatrix}.$$

Unknown matrices in Eq. (A3) are also defined as follows

$$\mathbf{P}_1 \mathbf{A}'_{ni} = \mathbf{P}_2 \mathbf{B}'_{ni} + \mathbf{P}_3 \tag{A.3}$$

$$P_1(j,m) = I_n(\lambda_m b) \times \tilde{\lambda}_{mj},$$

$$P_2(j,m) = \left( I_n(\lambda_m b) - \frac{I'_n(\lambda_m a)}{K'_n(\lambda_m a)} K_n(\lambda_m b) \right) \times \tilde{\lambda}_{mj}$$

$$+ \left( I'_n(\lambda_m b) - \frac{I'_n(\lambda_m a)}{K'_n(\lambda_m a)} K'_n(\lambda_m b) \right) \times \hbar_{mj}, \tag{A.4}$$

$$P_3(j) = \sum_{m=1}^{\infty} \left[ \frac{K'_n(\lambda_m b)}{K'_n(\lambda_m a)} \times \tilde{\lambda}_{mj} + \frac{K_n(\lambda_m b)}{K'_n(\lambda_m a)} \times \hbar_{mj} \right]$$

$$\times \frac{2}{H} \int_0^H \psi_i(x) \cos(\lambda_m x) \, dx,$$

**Appendix B**: the unknown matrices in Eq. (58).

In this appendix, the unknown matrices in Eq. (58) are given in detail. Unknown matrices in Eq. (B1) are as follows

$$\mathbf{Q}_1 \mathbf{A}_{sk} = \mathbf{Q}_2 \mathbf{B}_{sk} + \mathbf{Q}_3 \mathbf{S} \tag{B.1}$$

$$Q_1(j,f) = I'_n(\lambda_f b) \cos(\lambda_f x_j), \quad Q_3(j) = 0,$$

$$Q_2(j,f) = \left[ I'_n(\lambda_f b) - \frac{I'_n(\lambda_f a)}{K'_n(\lambda_f a)} K'_n(\lambda_f b) \right] \tag{B.2}$$

$$\times \cos(\lambda_f x_j), \quad \text{on} \quad \gamma$$

and

$$Q_1(j,f) = I'_n(\lambda_f b) \cos(\lambda_f x_j), \quad Q_2(j,f) = 0,$$
$$Q_3(j) = \Lambda_k(x_j), \quad \text{on} \quad \gamma' \tag{B.3}$$

$$\mathbf{A}_{sk} = \begin{Bmatrix} A_{1sk} \\ \vdots \\ A_{fsk} \end{Bmatrix}, \quad \mathbf{B}_{sk} = \begin{Bmatrix} B_{1sk} \\ \vdots \\ B_{fsk} \end{Bmatrix} \tag{B.4}$$

$$x_j = (j-1)\frac{H}{N}, \quad j = 1, \cdots, N$$

where $x_j$ are selected as points used in the collocation method. Moreover, unknown matrices in Eq. (B.5) are as follows

$$\mathbf{R}_1 \mathbf{B}_{sk} = \mathbf{R}_2 \mathbf{A}_{sk} + \mathbf{R}_3 \tag{B.5}$$

in which sub-matrices on the boundary $\gamma$ are as follows

$$R_1(j,f) = \left( I_n(\lambda_f b) - \frac{I'_n(\lambda_f a)}{K'_n(\lambda_f a)} K_n(\lambda_f b) \right)$$
$$\times \cos(\lambda_f x_j)$$
$$R_2(j,f) = I_n(\lambda_f b) \cos(\lambda_f x_j), \tag{B.6}$$
$$R_3(j) = 0,$$

and on the boundary $\gamma'$

$$R_1(j,f) = \left( I'_n(\lambda_f b) - \frac{I'_n(\lambda_f a)}{K'_n(\lambda_f a)} K'_n(\lambda_s b) \right)$$
$$\times \cos(\lambda_f x_j) \tag{B.7}$$
$$R_2(j,f) = 0, \quad R_3(j) = \Lambda_k(x_j)$$

**Appendix C**: the stiffness and the mass matrices

In this appendix, the stiffness and the mass matrices of the container and the baffle are given in detail. The stiffness matrix of the baffle is given by

$$\mathbf{K}_{S2} = \begin{bmatrix} \mathbf{K}^{11} & \mathbf{K}^{12} & \mathbf{K}^{13} \\ \mathbf{K}^{21} & \mathbf{K}^{22} & \mathbf{K}^{23} \\ \mathbf{K}^{31} & \mathbf{K}^{32} & \mathbf{K}^{33} \end{bmatrix} \tag{C.1}$$

where the elements of sub-matrices $\mathbf{K}^{kl}, (k,l=1,2,3)$ are given by Eq. (C.2)

$$\mathbf{K}^{11}_{injs} = \frac{A_s}{\alpha_i \alpha_j} \left[ b\varepsilon_c \int_{L-h}^{L} \frac{\partial^2 \Lambda_i}{\partial x^2} \frac{\partial^2 \Lambda_j}{\partial x^2} dx + \right.$$

$$\left. (1-\vartheta) \frac{ns\varepsilon_s}{2b} \int_{L-h}^{L} \frac{\partial \Lambda_i}{\partial x} \frac{\partial \Lambda_j}{\partial x} dx \right], \quad i,j = 1, \cdots M$$

in which $\varepsilon_c = \int_0^{2\pi} \cos(n\theta) \cos(s\theta) d\theta$

and $\varepsilon_s = \int_0^{2\pi} \sin(n\theta) \sin(s\theta) d\theta$ \tag{C.2}

$$\mathbf{K}^{12}_{injs} = \frac{A_s}{2\alpha_i} \left[ 2\vartheta n\varepsilon_c \int_{L-h}^{L} \frac{\partial^2 \Lambda_i}{\partial x^2} \Lambda_j \, dx \right.$$

$$\left. - (1-\vartheta)s\varepsilon_s \int_{L-h}^{L} \frac{\partial \Lambda_i}{\partial x} \frac{\partial \Lambda_j}{\partial x} dx \right]$$

$$\mathbf{K}_{injs}^{13} = \frac{A_s \vartheta \varepsilon_c}{\alpha_i} \int_{L-h}^{L} \frac{\partial^2 \Lambda_i}{\partial x^2} \Lambda_j \, dx$$

$$\mathbf{K}_{injs}^{22} = \frac{(1-\vartheta)\varepsilon_s}{2}(A_s b + \frac{D_s}{b}) \times$$
$$\int_{L-h}^{L} \frac{\partial \Lambda_i}{\partial x} \frac{\partial \Lambda_j}{\partial x} \, dx + \frac{n s \varepsilon_c}{b}(\frac{D_s}{b^2} + A_s) \int_{L-h}^{L} \Lambda_i \Lambda_j \, dx,$$

$$\mathbf{K}_{injs}^{23} = \frac{n\varepsilon_c}{b}(A_s + \frac{D_s s^2}{b^2}) \int_{L-h}^{L} \Lambda_i \Lambda_j \, dx +$$
$$D_s \left[ \frac{(1-\vartheta)s\varepsilon_s}{b} \int_{L-h}^{L} \frac{\partial \Lambda_i}{\partial x} \frac{\partial \Lambda_j}{\partial x} \, dx \right.$$
$$\left. - \frac{\vartheta n \varepsilon_c}{b} \int_{L-h}^{L} \frac{\partial^2 \Lambda_i}{\partial x^2} \Lambda_j \, dx \right],$$

$$\mathbf{K}_{injs}^{33} = \frac{A_s \varepsilon_c}{b} \int_{L-h}^{L} \Lambda_i \Lambda_j \, dx +$$
$$D_s \left[ b\varepsilon_c \int_{L-h}^{L} \frac{\partial^2 \Lambda_i}{\partial x^2} \frac{\partial^2 \Lambda_j}{\partial x^2} \, dx \right.$$
$$-2\vartheta \frac{s^2 \varepsilon_c}{b} \int_{L-h}^{L} \frac{\partial^2 \Lambda_i}{\partial x^2} \Lambda_j \, dx + \frac{(ns)^2 \varepsilon_c}{b^3} \times$$
$$\left. \int_{L-h}^{L} \Lambda_i \Lambda_j \, dx + 2 n s \varepsilon_s \frac{1-\vartheta}{b} \int_{L-h}^{L} \frac{\partial \Lambda_i}{\partial x} \frac{\partial \Lambda_j}{\partial x} \, dx \right]$$

The mass matrix of the baffle is given by

$$\mathbf{M}_{S2} = \begin{bmatrix} \mathbf{M}^{11} & 0 & 0 \\ 0 & \mathbf{M}^{22} & 0 \\ 0 & 0 & \mathbf{M}^{33} \end{bmatrix} \quad \text{(C.3)}$$

where the elements of sub-matrices $\mathbf{M}^{kk}$, ($k=1,2,3$) are given by

$$\mathbf{M}_{injs}^{11} = b\rho_s t_2 \varepsilon_c \int_{L-h}^{L} \frac{1}{\alpha_i \alpha_j} \frac{\partial \Lambda_i}{\partial x} \frac{\partial \Lambda_j}{\partial x} \, dx,$$
$$\mathbf{M}_{injs}^{22} = \mathbf{M}_{injs}^{33} = b\rho_s t_2 \varepsilon_c \int_{L-h}^{L} \Lambda_i(x) \Lambda_j(x) \, dx, \quad \text{(C.4)}$$
$$i,j = 1 \cdots M$$

The partitioned stiffness matrix $\mathbf{K}_{S1}$ associated with the container is written as follows

$$\mathbf{K}_{S1} = \begin{bmatrix} [\mathbf{K}_{11}] & [\mathbf{K}_{12}] & [\mathbf{K}_{13}] \\ [\mathbf{K}_{12}]^T & [\mathbf{K}_{22}] & [\mathbf{K}_{23}] \\ [\mathbf{K}_{13}]^T & [\mathbf{K}_{23}]^T & [\mathbf{K}_{33}] \end{bmatrix}, \quad \text{(C.5)}$$

where the elements of sub-matrices $[\mathbf{K}_{ij}]$, ($i,j=1,2,3$) are given by Eq. (C.6)

$$[\mathbf{K}_{11}]_{injs} = \frac{A_s \varepsilon_c}{\alpha_i \alpha_j} \left[ a \int_0^L \frac{\partial^2 \psi_i}{\partial x^2} \frac{\partial^2 \psi_j}{\partial x^2} \, dx \right.$$
$$\left. + (1-\vartheta) \frac{ns}{2a} \int_0^L \frac{\partial \psi_i}{\partial x} \frac{\partial \psi_j}{\partial x} \, dx \right], \quad \text{(C.6)}$$

$$[\mathbf{K}_{12}]_{injs} = \frac{A_s}{2\alpha_i} \left[ -2\vartheta n \varepsilon_c \int_0^L \frac{\partial^2 \psi_i}{\partial x^2} \psi_j \, dx \right.$$
$$\left. + (1-\vartheta) s \varepsilon_s \int_0^L \frac{\partial \psi_i}{\partial x} \frac{\partial \psi_j}{\partial x} \, dx \right],$$

$$[\mathbf{K}_{13}]_{injs} = -\frac{A_s \vartheta \varepsilon_c}{\alpha_i} \int_0^L \frac{\partial^2 \psi_i}{\partial x^2} \psi_j \, dx$$

$$[\mathbf{K}_{22}]_{injs} = \frac{1-\vartheta}{2}\left( A_s a + \frac{D_s}{a} \right) \varepsilon_s \int_0^L \frac{\partial \psi_i}{\partial x} \frac{\partial \psi_j}{\partial x} \, dx$$
$$+ \frac{n\varepsilon_c}{a}\left( \frac{D_s}{a^2} + A_s \right) \int_0^L \psi_i \psi_j \, dx,$$

$$[\mathbf{K}_{23}]_{injs} = \frac{n\varepsilon_c}{a}\left( A_s + \frac{s^2 D_s}{a^2} \right) \int_0^L \psi_i \psi_j \, dx +$$
$$\frac{D_s}{a}\left[ (1-\vartheta) s \varepsilon_s \int_0^L \frac{\partial \psi_i}{\partial x} \frac{\partial \psi_j}{\partial x} \, dx \right.$$
$$\left. - \vartheta n \varepsilon_c \int_0^L \frac{\partial^2 \psi_i}{\partial x^2} \psi_j \, dx \right],$$

$$[\mathbf{K}_{33}]_{injs} = \frac{A_s \varepsilon_c}{a} \int_0^L \psi_i \psi_j \, dx +$$
$$D_s \left[ a\varepsilon_c \int_0^L \frac{\partial^2 \psi_i}{\partial x^2} \frac{\partial^2 \psi_j}{\partial x^2} \, dx + 2\vartheta \varepsilon_c \frac{s^2}{a} \int_0^L \frac{\partial^2 \psi_i}{\partial x^2} \psi_j \, dx \right.$$
$$\left. + \frac{(ns)^2 \varepsilon_c}{a^3} \int_0^L \psi_i \psi_j \, dx + 2 n s \varepsilon_s \frac{1-\vartheta}{a} \int_0^L \frac{\partial \psi_i}{\partial x} \frac{\partial \psi_j}{\partial x} \, dx \right]$$

The mass matrix of the container is written

$$\mathbf{M}_{S1} = \begin{bmatrix} [\mathbf{M}_{11}] & [0] & [0] \\ [0] & [\mathbf{M}_{22}] & [0] \\ [0] & [0] & [\mathbf{M}_{33}] \end{bmatrix}, \quad \text{(C.7)}$$

where the elements of sub-matrices $[\mathbf{M}_{ii}]$, ($i=1,2,3$) are given by

$$[\mathbf{M}_{11}]_{injs} = a\rho_s t_1 \varepsilon_c \int_0^L \frac{\partial \psi_i}{\alpha_i \partial x} \frac{\partial \psi_j}{\alpha_j \partial x} \, dx,$$
$$[\mathbf{M}_{22}]_{injs} = [\mathbf{M}_{33}]_{injs} = a\rho_s t_1 \varepsilon_c \int_0^L \psi_i(x) \psi_j(x) \, dx \quad \text{(C.8)}$$